\documentclass[12pt]{iopart}

\usepackage{bm}
\usepackage{graphicx}
\usepackage{url}
\usepackage{hyperref}
\begin{document}

\title[Li et al]{Impurity effects on ion temperature gradient and trapped electron modes in reversed-field pinch plasmas of hollow density profiles}
\author{Jingchun Li$^{1}$, Songfen Liu$^2$, Yilong Zhang$^{2}$, Jiaqi Dong$^{3,4,5}$ and Wei Kong$^{6}$}

\address{$^1$Department of Earth and Space Sciences, Southern University of Science and Technology, 518055 Shenzhen, Guangdong, People's Republic of China\\
$^2$School of Physics, Nankai University, Tianjin 300071, People's Republic of China\\
$^3$Southwestern Institute of Physics, Chengdu 610041, People's Republic of China\\
$^4$Hebei Key Laboratory of Compact Fusion, Langfang 065001, China \\
$^5$ENN Science and Technology Development Co., Ltd., Langfang 065001, China \\
$^6$College of Science, Civial Aviation University of China, Tianjin 300300, People's Republic of China}
%$^4$ENN Group, Langfang 065001, People¡¯s Republic of China\\
%$^5$School of Physics, Nankai University, Tianjin 300071, People's Republic of China}
\ead{jingchunli@pku.edu.cn}
\vspace{10pt}
\begin{indented}
\item[]August 2021
\end{indented}

\begin{abstract}
The drift wave in the presence of impurity ions was investigated numerically in reversed field pinch (RFP) plasmas, using the gyrokinetic integral eigenmode equation. It was found that in RFP plasmas with hollow density profiles, an increase in $k_\theta \rho_s$ increases the growth rate of the ion temperature gradient (ITG). Comparing the results of regular and hollow plasma density profile shows that the ITG mode under the hollow density profile is much harder to be excited. For the impurities' effects, when the impurities' density gradient is opposite to the primary ions, namely when $L_{ez}$ is negative, impurities could enhance the instability. On the contrary, when $L_{ez}$ is positive, the instability is stabilized. Regarding the trapped electron mode (TEMs), the growth rate under the plasma with hollow density profile remained minor than that for the standard density gradient. There exists a threshold of $L_{ez}$. When $L_{ez}$ is less than this value, impurity destabilizes TEMs, while as $L_{ez}$ is greater than this, impurity stabilizes TEMs. The effects of $L_{ez}$ on TEM also depend on both the plasma density gradient and the impurity species. In addition, the influence of collisionality on TEMs was also studied.

\end{abstract}

%
% Uncomment for keywords
%\vspace{2pc}
%\noindent{\it Keywords}: XXXXXX, YYYYYYYY, ZZZZZZZZZ
%
% Uncomment for Submitted to journal title message
%\submitto{\JPA}
%
% Uncomment if a separate title page is required
%\maketitle
%
% For two-column output uncomment the next line and choose [10pt] rather than [12pt] in the \documentclass declaration
%\ioptwocol
%
\section{\label{sec:level1}INTRODUCTION}

%\section{Introduction}
In the toroidal plasma, the distribution function of the plasma in the velocity space deviates from the Maxwell distribution, and the instability that develops only on the microscopic scale is called microscopic (in the velocity space) instability. As the plasma develops into a nonlinear saturation state, plasma turbulence will be generated, which causes macroscopic turbulent transport [1,2,3,4,5]. Among these perturbations, the micro-instability driven by the ion temperature gradient (ITG) is considered to be the most likely cause of the thermal turbulence transport of the ions, and the trapped electron mode (TEM) may be the main factor causing the abnormal transport [6,7,8]. Drift-type microturbulence is a rather new research subject in the reversed-field pinch (RFP) community although it was a hot topic in the tokamak community about four decades ago [2]. This is because considerable efforts have been devoted to the magnetohydrodynamics (MHD) macroturbulence. Nevertheless, the interest in the study of this type of instabilities in RFP plasmas is increasing because higher performances have been obtained in RFP plasmas, and one should consider the role possibly played by the drift wave turbulence for further improving the confinement in RFP plasmas.
In the actual operation of a toroidal plasma, particle injection often leads to an abnormal (hollow) distribution of the plasma density [9,10], and the generation of impurities is also unavoidable [11,12]. Therefore, understanding impurities and their influence on the microturbulence in RFP plasmas with both normal and hollow density profiles is one of the critical challenges that magnetically confined fusion plasmas face[13,14].

Currently, microscopic instabilities have been extensively studied experimentally. In the RFX-mod, the axisymmetric and helical RFP configuration with different $q$(a) states (reduced from 4 to 1.2) have been studied [15]. In the RFX-mod, two transmission parameters, namely the diffusion coefficient and the convection velocity, have been determined, and previous works also confirmed the presence of external velocity barriers for light impurities [16]. In terms of particle injection, Ref. [17] determined the drift of the deposited material in the non-uniform magnetic field, thus providing the possibility to effectively inject the plasma from the high field side. The particle injection experiment undertaken by HL-2A tokamak achieved enhanced confinement, indicating that the improvement of confinement is related to a reduction in electron heat transport [18]. Repetitive particle injection experiments performed on a large helical device (LHD) showed an improved energy confinement, which could be maintained over time [19]. In particle injection experiments in RFP plasmas, the plasma evolution after particle injection was also analyzed [13].

The microinstability related to the hollow density profile in a toroidal plasma has been theoretically investigated in recent years [20-21]. Under the slab configuration with magnetic shear, the instability driven by the electron temperature gradient in a plasma with a slight hollow density profile was studied using the gyrokinetic integral eigenvalue equation, showing the suppression effect of shear flow in transport [22]. Using the gyrokinetic eigenmode equation, the ITG mode was modeled numerically in the presence of impurity ions and trapped electrons (TEs) in the tokamak plasma with a hollow density distribution. It was found that, in the hollow density plasma, the increase of ITG results in an increase of the ITG growth rate and frequency, and that the density gradient plays an important role in the ITG mode [23].

For the RFP configurations, when considering impurities, the RFX-mod device usually exhibits a hollow carbon/oxygen distribution, which peaks in the edge region. Ref. [24] calculated the ITG instability and gyrokinetic results of turbulence, and described the role of impurities in the ITG mode instability. Ref. [25] used a gyro-kinetic integral equation containing trapped electrons and impurity ions to study the electrostatic mode driven by ITG in RFP plasma. It was proved that, when the density gradient is opposite to the primary ion, the impurity ions cause an increase in the ITG instability, and the trapped electrons also have a destabilizing effect on the ITG mode. Ref. [26] studied the stability threshold of the ITG mode through the linear gyrokinetic theory, showing that the temperature slope required to excite the ITG instability is much steeper than that of tokamak. In addition, Ref. [27-28] considered TEs and all-ion kinetic effects, and found that the role of TEs becomes important in RFP only under a very steep density/temperature gradient. Compared with the tokamak plasma, the instability of TEM in the RFP plasma requires a much larger density/temperature gradient, and the $k_\theta \rho_s$ spectrum is much narrower.

To summarize, the impurities and their influence on microturbulence in tokamaks with a hollow density have been clarified. However, these effects have not been investigated in RFP plasmas; in particular, the impurity effect on the TEMs remains unclear. In this work, the drift wave instability is studied under RFP configurations. We considered both collisionless and collisional RFP plasmas with standard and hollow density profiles. By comparing the results of the typical plasma density profile ($\varepsilon_{n} > 0$) and the hollow plasma density profile ($\varepsilon_{n}< 0$), it is shown that the growth rate of the ITG instability in the plasma with the recessed density profile is lower and is harder to excite. Moreover, when $L_{ez}$ is positive, the impurities can stabilize the instability, while when $L_{ez}$ is negative, the impurities enhance the instability. Regarding the TEMs, the growth rate for the plasma with a hollow density profile remains smaller than that for the plasma with a standard density profile. Besides, the collisionality can stabilize the TEMs.

The remainder of this article is organized as follows. The integral eigenvalue equations are introduced in Section II. The numerical results of the drift wave in RFP plasmas with a hollow density profile are presented and analyzed in Sections III. Finally, the conclusions are drawn in Section IV.

\section{PHYSICAL MODEL}

Considering the effect of trapped electron and impurity ions, the quasi-neutral condition can be written as
%\begin{equation}
%	\tilde{n}_{i}+Z\tilde{n}_{z}=(1-f_{te})\tilde{n}_{pe}+f_{te}\tilde{n}_{te}
%\end{equation}
\begin{equation}
	\tilde{n}_{i}+Z\tilde{n}_{z}=(1-f_{te})\tilde{n}_{pe}+\tilde{n}_{te}
\end{equation}

Here, $\tilde{n}_{i},\tilde{n}_{z},\tilde{n}_{pe},\tilde{n}_{te}$ are the perturbed densities of the main ions, impurity ions, passing electrons, and trapped electrons, respectively.

In the balloon representation, it is possible to write the differential equation which is satisfied by the particle non-adiabatic response:
%\begin{equation}
%	\left(\omega-\omega_{Dj}+iv_\parallel\boldsymbol{b}\cdot\nabla\right)h_j %\notag
%	=\left(\omega-\omega_{\ast Tj}\right)J_0\left(\frac{k_\perp v_\perp}{\omega_{cj}}\right)\frac{q_jF_{Mj}}{T_j}\phi %\label{wffc}
%\end{equation}

\begin{equation}
	\left [ \frac{v_{||}}{Rq_j \alpha}\partial_x-i(\omega-\omega_{Dj}) \right]h_j=-i(\omega-\omega_{\ast Tj})J_0(\mu)F_{mj}\phi(x)%\right. %\label{wffc}
\end{equation}
where, $\omega,\omega_{Dj},\omega_{\ast Tj}$ are the mode eigenfrequency, the magnetic drift frequency, and the diamagnetic drift frequency caused by the pressure gradient. Their specific expressions are:  % under the  $s-\alpha$  local equilibrium model are:
\begin{eqnarray}
\omega_{Dj}=\hat k_\theta \rho_j v_{tj}[(\hat v^2_\perp/2)/L_B-(\varepsilon^2/q^2_j\alpha^2)\hat v^2_{||}/r]\\
\omega_{\ast Tj}=\omega_{\ast j}\left[1+\eta_j\left(\hat v^2_\perp+\hat v^2_{||} -\frac{3}{2}\right)\right]
\end{eqnarray}
where $\omega_{\ast j}=\hat k_\theta \rho_i v_{tj}/2L_n$, $\varepsilon=r/R$, with $R$ being the major radius of the
torus. $j = e, i, z$ represent electrons, main ions, and impurity ions, respectively. Velocity and length are normalized to: $v_{ti}$ and $\frac{\sqrt{\tau_i}v_{ti}}{\omega_{ci}}$ respectively; $\hat v_\perp$ and $\hat v_{||}$ being
the velocity components perpendicular and parallel to the
magnetic field, respectively; $F_{mj}$ is a Maxwellian equilibrium distribution function; $h_j$ can be solved from the differential equation (2). Using formula
$\tilde{n}_j=\int f_jd^3v$, the perturbation density of passing electrons, hydrogen ions, and impurity ions can be obtained. Thus, by substituting these quantities into the quasi-neutral condition equation, the integral eigenvalue equation can be achieved. In the RFP configuration, the toroidal magnetic field reverses in the edge
of plasma during the plasmas relaxation, and has the
same order of magnitude as that of the poloidal magnetic field. The
formula of poloidal and toroidal magnetic field in RFPs can be found in Ref.[25].

The integral eigenvalue equation is written as follows:

\begin{eqnarray}
[1+\tau_i(1-\sum_z f_{z})+\sum_z z_z\tau_{z}f_{z}]\hat{\phi}(k)=-\widetilde{n}_{te}+\nonumber\\
\int_{-\infty}^{-\infty}\frac{dk'}{\sqrt{2\pi}}\tilde{K}(k,k')\hat{\phi}(k')\ .
\end{eqnarray}
The kernel $\tilde{K}(k,k')$ obtained from the gyrokinetic equation for all types of ion species can also be found in Ref.[25]. $\tau_i=T_e/T_i$ and $f_z$ are distribution function of the particles. Note that these parameters are not independent from each other. From the quasi-neutral condition, it can be derived that:
\[ L_{ei}=\frac{1-f_zL_{ez}}{1-f_z}. \]
%	\[ \eta_z=\frac{\eta_i\left( L_{nz}-f_zL_{ne} \right)}{L_{ne}\left( 1-f_z \right)} \]
Where $L_{ez} = L_{ne}/L_{nz}$, and $L_{n}=-(d ln(n) /d r)^{-1}$ is the gradient scale length.

The trapped electron response in the ballooning representation is presented as:

\begin{eqnarray}
 \tilde{n}_{te}=-\frac{en_{0e}}{T_e}\sqrt{\frac{2\varepsilon}{\pi}}\int_0^{+\infty}dt\sqrt{t}e^{-t}\int_0^1\frac{\omega-\omega_\ast^e}{\omega-\bar{\omega}_d^e+i\nu_{e\mathrm{eff}}\cdot t^{-\frac{3}{2}}}\cdot  \nonumber  \\%\notag \\
 \frac{d\kappa^2}{4F(\kappa)}\cdot\sum_{j=-\infty}^{+\infty}g\left( \theta-2\pi j,\kappa \right)\int_{-\infty}^{+\infty}d\theta'g(\theta',\kappa)\phi\left( \theta'-2\pi j \right)
\end{eqnarray}
where
\[ g(\eta,\kappa)=\int_{-\theta_r}^{+\theta_r}\frac{\delta(\eta-\theta')d\theta'}{\sqrt{\kappa^2-\sin^2\frac{\theta'}{2}}} \],
\[ \kappa=\sin^2\frac{\theta_r}{2} , \varepsilon=\frac{r}{R} , \nu_{e\mathrm{eff}}=\frac{\nu_{ei}}{\varepsilon} \],

\[ \bar{\omega}_d^e
%=\omega_{\ast e}\varepsilon_ntG(s,\kappa)
=\omega_{\ast e}\varepsilon_nt\left[ \frac{2F(\kappa)}{K(\kappa)}-1+4s\left( \frac{F(\kappa)}{K(\kappa)}-1+\kappa^2 \right) \right] \]
\[ \omega_\ast^e=\omega_{\ast e}\left[ 1+\eta_e\left( t-\frac{3}{2} \right) \right] \]
\[ K(\kappa)=\int_0^{\frac{\pi}{2}}\frac{d\kappa}{\sqrt{1-\kappa^2\sin^2\theta}} \]
\[ F(\kappa)=\int_0^{\frac{\pi}{2}}d\kappa \sqrt{1-\kappa^2\sin^2\theta} \]

In the above equations, $\kappa$ is the pitch angle, $\theta_r$ is the angle corresponding to the turning point of the trapped electron, and $\bar{\omega}_d^e$ is the precession drift frequency of the TEs. The last two equations are elliptic functions.

In the following section, the gyrokinetics eigenvalue code HD7 is used to solve equation (5), and the numerical results are discussed.

\section{SIMULATION RESULTS}

We first study the ITG mode in RFP plasmas. Namely, we do not consider the first term on the right side of the equation (5). The default parameter we utilized are $\hat s  = 1.0, \tau_i =\tau_z= 1.33, \eta_z=16, k_\theta \rho_s = 0.447, \varepsilon = 0.0, q = 0.15$ unless otherwise stated.

\subsection{ITG modes}
Firstly, the effects of the ion temperature gradient $R/L_{Ti}$ and density gradient $R/L_n$ on the ITG modes are discussed because they are generally believed to provide the driving force for the ITG modes. The normalized growth rates of the pure ITG modes as a function of $R/L_{Ti}$ and $R/L_n$ are illustrated in Figs. 1(a) and 1(b), respectively. We can see from Fig. 1(a) that the growth rate of the ITG modes increases with the increase in $R/L_{Ti}$. The larger the absolute value of $R/L_n$ is, the greater the growth rate of the ITG modes is. Fig. 1(b) shows that the smaller the $R/L_{Ti}$, the smaller the ITG growth rates. At the same time, when the absolute values of $R/L_n$ are equal, the growth rate corresponding to positive $R/L_n$ ratios is much larger. This indicates that the ITG mode in plasmas with a hollow density profile will be relatively difficult to excite in both the strong and weak density gradient cases compared with plasmas with typical density profiles. It is also possible to see that the growth rate is always monotonous when keeping the $R/L_n$ ratio constant. We can conclude that a high $R/L_{Ti}$ enhances the ITG instability, and it is easier to trigger the ITG mode in RFP plasmas with a standard density profile. %We could conclude that high $R/L_{Ti}$ enhances ITG instability, and it is easier to trigger the ITG mode in the RFP plasma with standard density profiles.

\begin{figure}
 \centering
  \includegraphics[width=8.5cm]{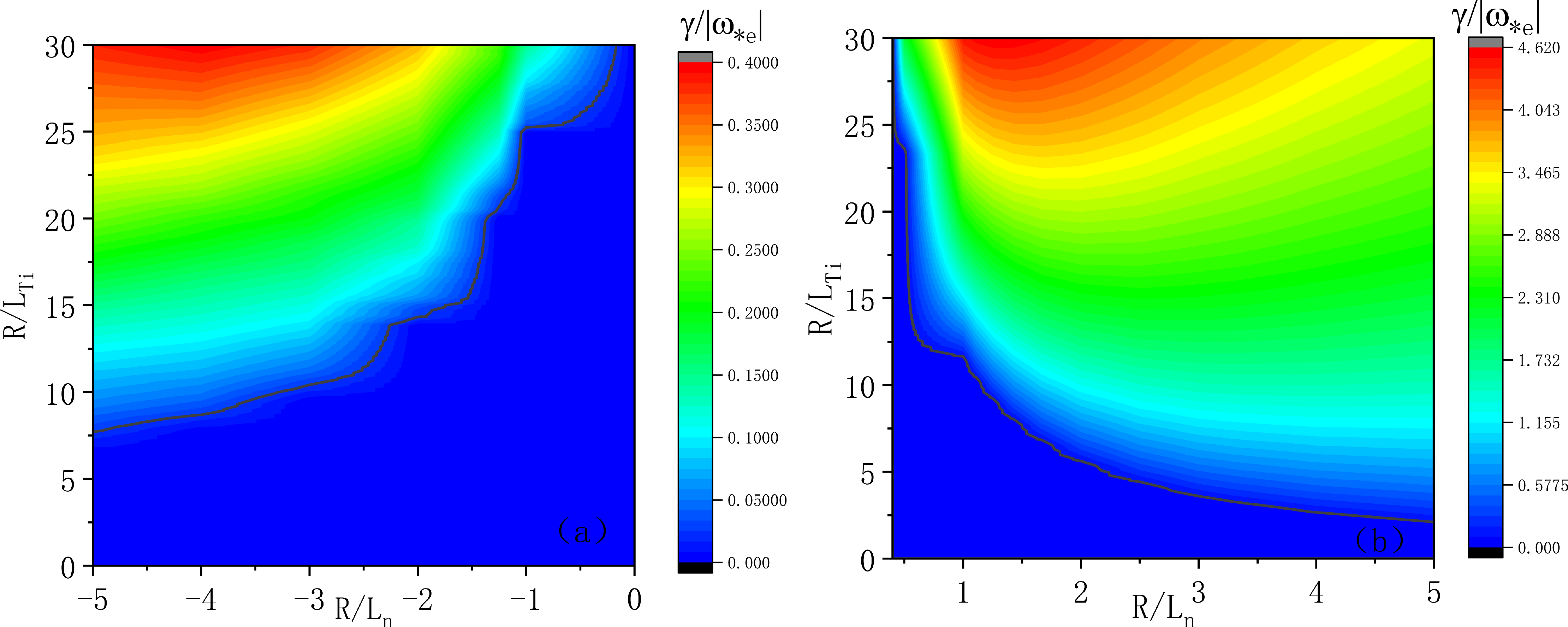}\\
  \caption{Contour plots of the normalized real frequency of the ITG mode in the plane of the normalized ion temperature gradient $R/L_{Ti}$ and density gradients $R/L_n$ in pure plasma(fz=0). (a) and (b) corresponds to the plasma with a hollow density profile, and a normal density profile, respectively. Other parameters are $\hat s  = 0.3, \tau_i = 1.33, k_\theta \rho_s = 0.447, \varepsilon = 0.0, q = 0.15$ .}\label{fig:fig1}
\end{figure}

%\subsection{Impurities effect}
The normalized real frequency and growth rate as a function of the charge concentration $f_z$ of fully ionized carbon impurities in plasmas with regular and hollow density profiles are represented in Fig. 2. Figs. 2(a) and 2(c) show that the ITG frequency and growth rate change monotonically with $f_z$. With positive values of $L_n$, it is clearly demonstrated that $L_{ez}$ = 2 has a stabilizing effect on the modes, while $L_{ez} = -6$ has a destabilizing impact on the modes. It is thus possible to conclude that when the impurity ion density gradient is opposite to that of the electrons ($L_{ez} < 0$), the ITG instability is enhanced. On the contrary, when $L_{ez} > 0$, the growth rate of the ITG instability is reduced. These findings are consistent with previous simulation results discussed in Refs. [23, 28]. Figs. 2(b) and 2(d) display the growth rate and real frequency of the ITG modes for the hollow density case. It can be seen that these results are very similar to those for the non-hollow density profile case when $L_{ez} < 0$; the ITG instability is reinforced. On the other hand, when $L_{ez} > 0$, the ITG growth rate first increases and then decreases with $f_z$. Furthermore, the greater $|L_{ez}|$, the greater the stabilizing/destabilizing effects. With a hollow density profile, the small growth rate of the ITG mode can still be seen in Fig. 2(b). Figs. 2(a) and 2(c) also reveal that the charge concentration can enhance the stabilizing/destabilizing effects of the carbon impurities.

\begin{figure}
 \centering
  \includegraphics[width=8.2cm]{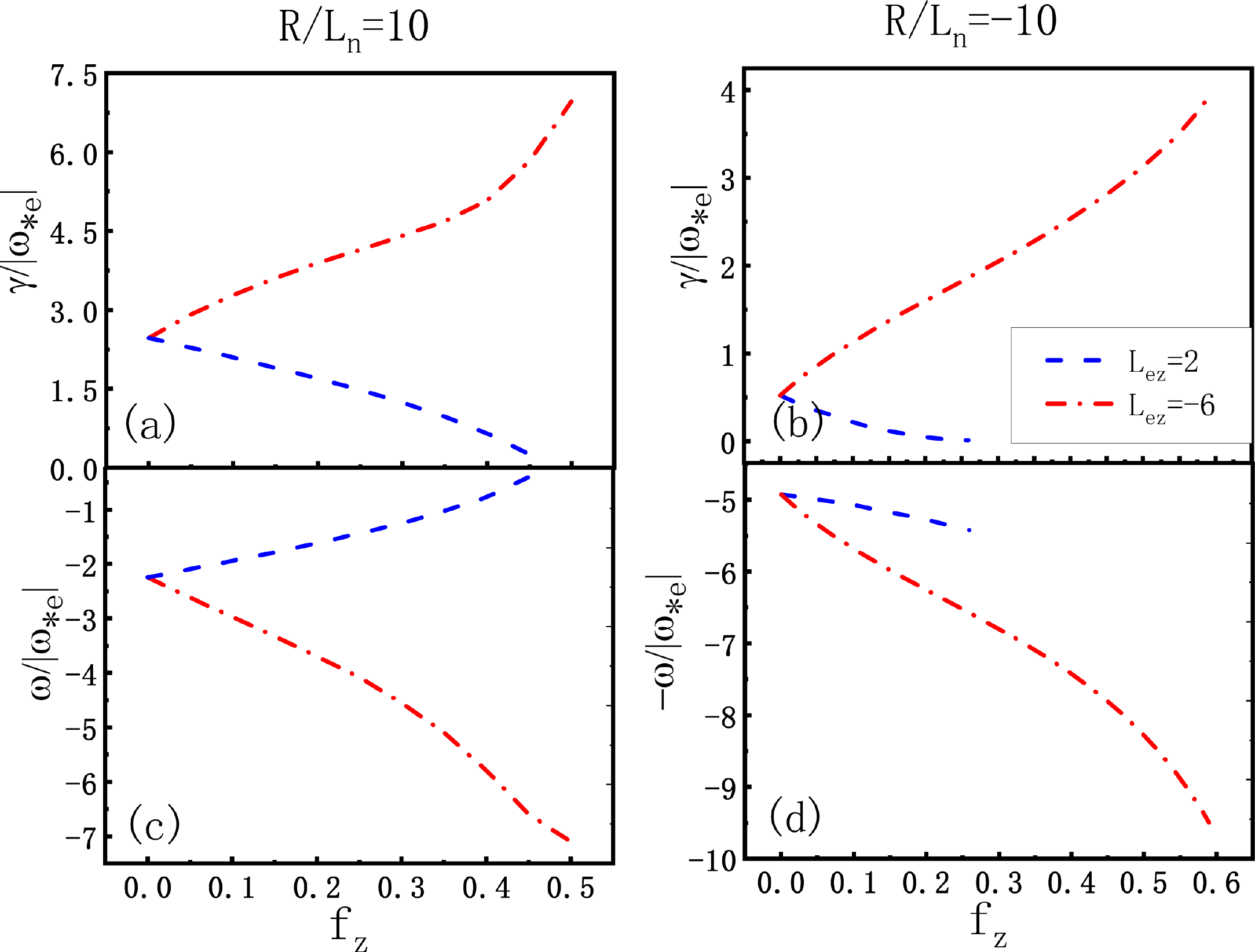}\\
  \caption{ Normalized real frequency and growth rate versus charge concentration $f_z$ of the fully ionized carbon impurity in plasmas with normal density profile (a)(c) and hollow density profile(b)(d). Other parameters are $\hat s  = 1.0, \tau_i =\tau_z= 1.33, \eta_z=16, k_\theta \rho_s = 0.447, \varepsilon = 0.0, q = 0.15$.}\label{fig:fig2}
\end{figure}

To further investigate the effect of $L_{ez}$ on the ITG modes, we scanned $L_{ez}$ for different density gradients. The normalized growth rate as a function of $L_{ez}$ is shown in Fig. 3 for different impurity charge concentrations of carbon. It shows that when $L_{ez}$ is positive, the impurities have a stabilizing effect. The larger the number of impurities, the stronger the stabilizing effect; on the other hand, when $L_{ez}$ is negative, the presence of impurities has a destabilizing effect. Besides, the influence of $L_{ez}$ on the ITG instability for plasmas with hollow and non-hollow density profiles is the same. The obvious difference is that in the case of the hollow density profile, the ITG growth rate is lower, which is consistent with Figs. 1 and 2. %The calculations demonstrate the ITG frequency increases monotonically and almost linearly with $L_{ez}$. Additionally, when $L_{ez}$ is positive/negative, the frequency increases/decreases monotonically with $f_z$.

\begin{figure}
 \centering
  \includegraphics[width=6.0cm]{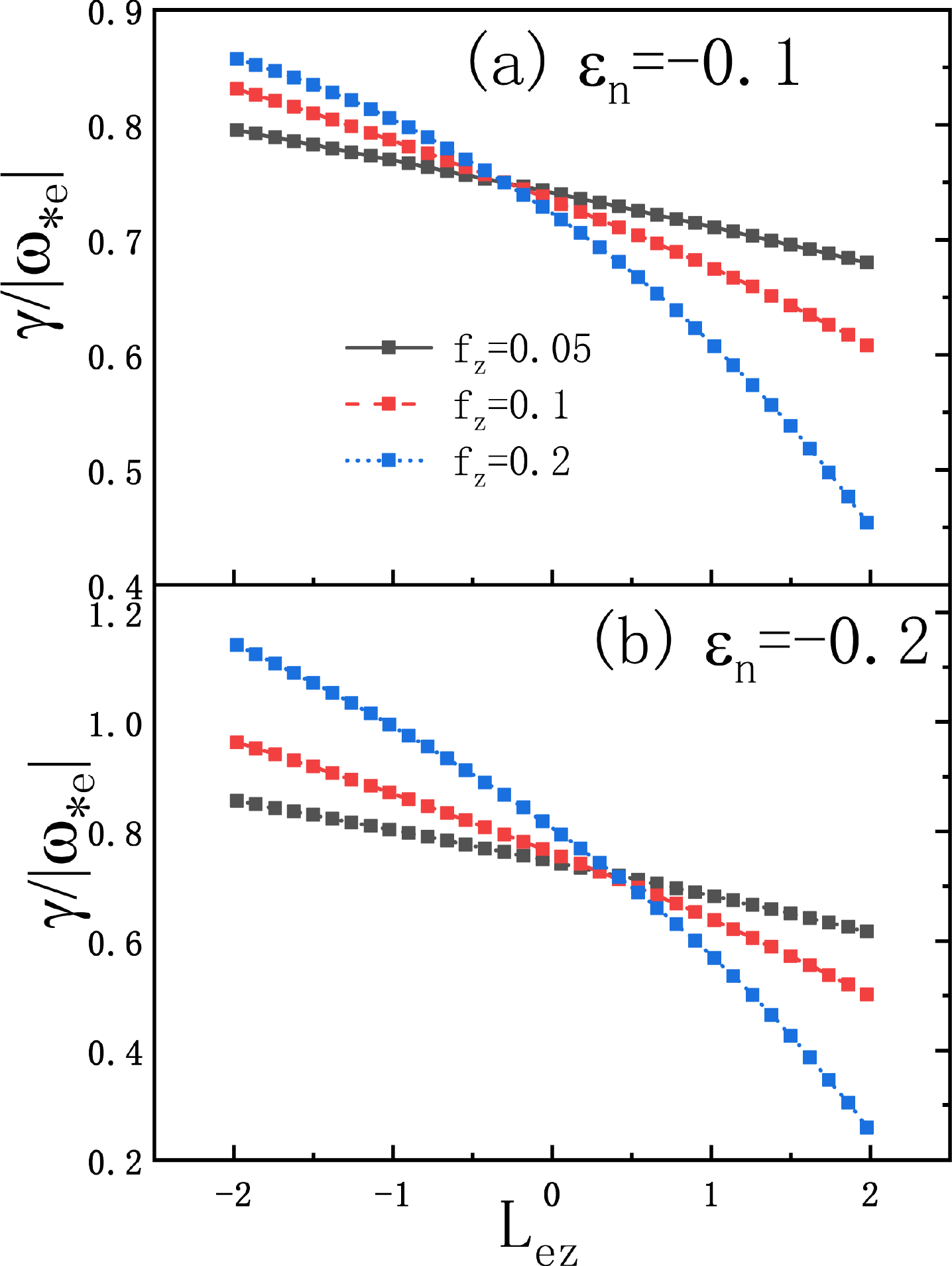}\\
  \caption{Normalized real frequency and growth rate versus $L_{ez}$ for different carbon ($C^{+6}$) impurity charge concentrations $f_z$. The other parameters are $\hat s  = 1.0, \tau_i =\tau_z= 1.33, \eta_z=16, k_\theta \rho_s = 0.447, \varepsilon = 0.0, q = 0.15, z = 6$ (carbon).}\label{fig:fig2}
\end{figure}

Since there are many impurity ions with different ionization states in RFP plasmas, one needs to investigate their impact individually. Here, we consider the influence of the $O^{+6}$, $C^{+8}$, and $W^{+8}$ impurities with varying ionization states. Fig. 4 shows the relationship between the ITG growth rate versus $L_{ez}$ with different ionized impurity contents. Figs. 4(a) and 4(b) correspond to the non-hollow density case, while Figs. 4(c) and 4(d) correspond to the hollow density case. We find that the growth rate of the ITG mode gradually decreases with the increase in the impurity mass. This trend can be seen for both weak and strong density gradients. Due to the extremely high mass number of tungsten, when $\varepsilon_n=-0.1$, this phenomenon is offset to a certain extent. Besides, Fig. 4 clearly demonstrates that the ITG growth rate decreases as $L_{ez}$ changes from negative to positive values.

We performed a further study on the impact of the tungsten impurities on the ITG. Considering ionized impurities ($W^{+6}$, $W^{+18}$, and $W^{+30}$) as impurity species, Fig. 5 depicts the normalized growth rate $\gamma/|\omega_{\star e}|$ as a function of $L_{ ez}$ for the hollow density profile case with different values of $\varepsilon_n$. One can see that the larger the impurity charge, the smaller the ITG growth rate, suggesting that compared with low-charge impurities, tungsten ions with a higher charge have a stronger destabilizing effect. By combining the above results, we can conclude that the impact of impurities on the ITG mode depends not only on the impurity density gradient and quality but also on the number of charges of the impurity. Most importantly, heavy impurities, such as tungsten, have a better stabilizing effect on the ITG mode.

\begin{figure}
 \centering
  \includegraphics[width=8.6cm]{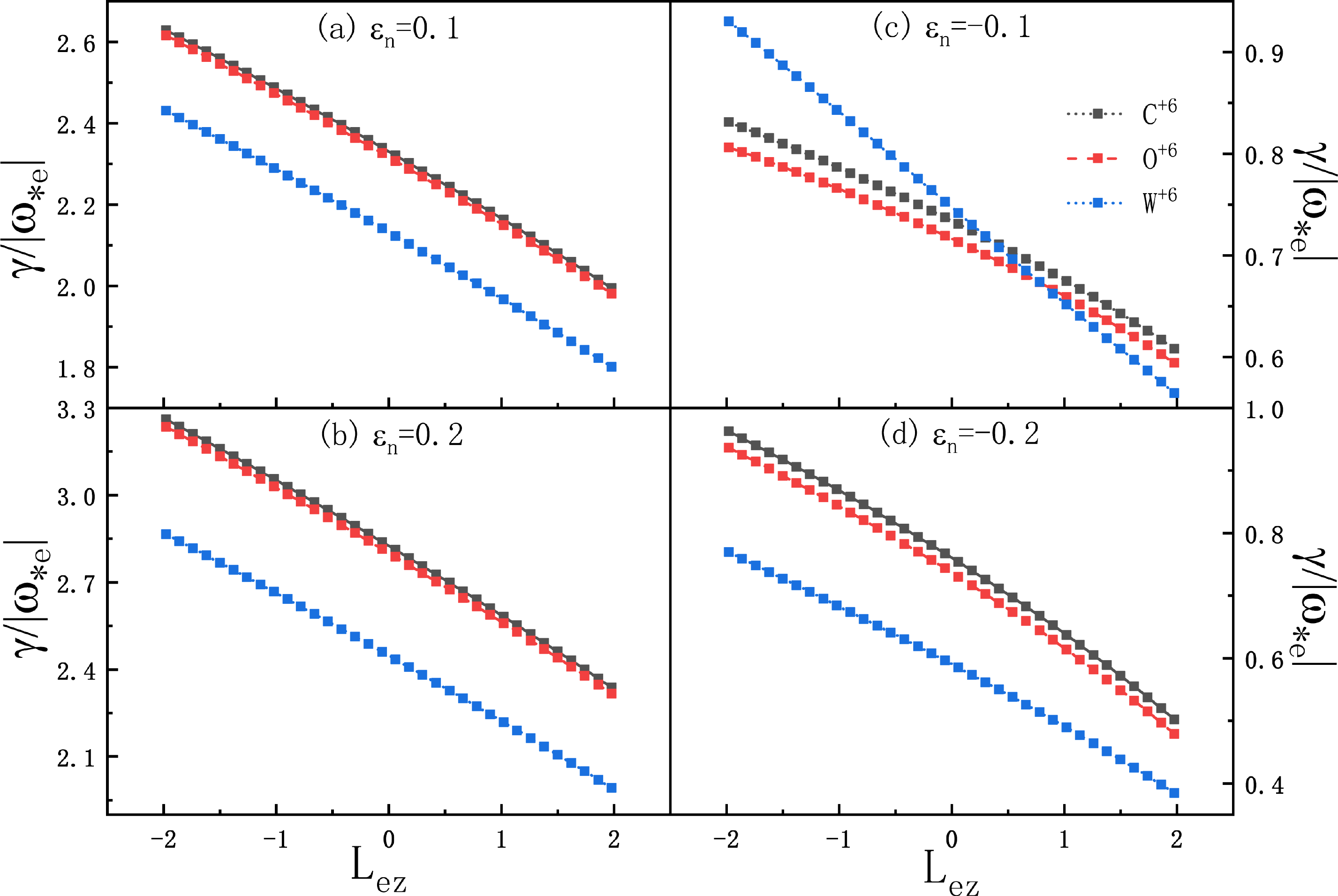}\\
  \caption{Normalized growth rate $\gamma/|\omega_{\star e}|$ and real frequency $\omega/|\omega_{\star e}|$ of ITG versus $L_{ez}$ under normal density profile (a)(c) and hollow density profile(b)(d) with different impurities($f_z=0.1$). Other parameters are the same as those used in figure 3.}\label{fig:fig3}
\end{figure}

\begin{figure}
 \centering
  \includegraphics[width=6.0cm]{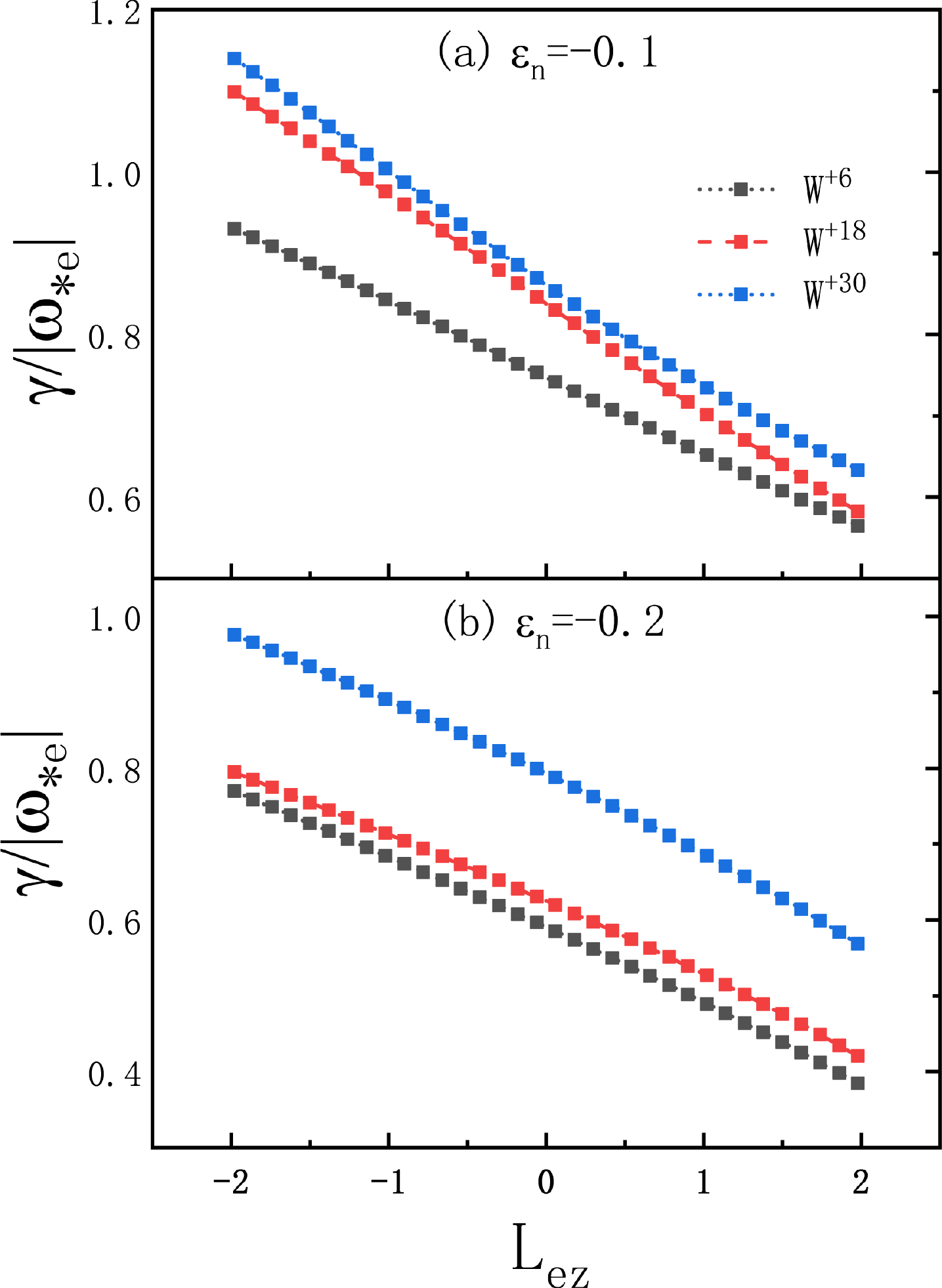}\\
  \caption{Normalized growth rate $\gamma/|\omega_{\star e}|$ versus $L_{ez}$ under hollow density profile with different $\varepsilon_n$,(a)$\varepsilon_n$=-0.1,(b)$\varepsilon_n$=-0.2. Ionized impurities ($W^{+6}$, $W^{+18}$, and $W^{+30}$) are taken as impurity species with $f_z=0.1$. Other parameters are the same as those used in figure 3.}\label{fig:fig5}
\end{figure}

Finally, we investigated the ITG stability threshold considering carbon impurities. Fig. 6 displays the temperature gradient threshold $\varepsilon_T$ for the ITG instability as a function of $\varepsilon_n$ for different $L_{ez}$ values for both hollow (Fig. 6(a)) and standard (Fig. 6(b)) density profiles. Figs. 6(a) and 6(c) suggest that the threshold decreases with increasing the absolute value of $\varepsilon_n$, and the larger $\varepsilon_n$ is, the more difficult it becomes for the ITG instability to be excited. However, the sign of $\varepsilon_n$ influences $\varepsilon_T$, and the $\varepsilon_T$ corresponding to the positive $\varepsilon_n$ is larger than the $\varepsilon_T$ corresponding to the negative one; that is, the modes with a positive $\varepsilon_n$ are more unstable. Impurities with a negative $L_{ez}$ can destabilize the mode and weaken the Landau damping effects of the primary ion in RFP plasmas. In addition, regardless of whether it is a hollow density or a normal density gradient, when impurities exist, and their density gradient is opposite to that of the plasma, the instability interval will increase. Overall, since the excitation range for the hollow density gradient is narrower than for the positive density gradient, the ITG in the hollow plasma is more difficult to trigger. These results are in agreement with cases in tokamaks [23]. In tokamak plasmas, ITG modes can be easily excited by ion temperature gradients in the negative density gradient range for relatively large density gradients.
In RFP plasmas, when the density is large, the temperature gradient interval of the ITG instability is relatively large. Thus, impurities with a negative $L_{ez}$ enhance the ITG modes in RFP plasmas. The instability interval for the hollow density case is smaller than that for the typical density case. Nevertheless, if the negative $L_{ez}$ and negative density gradients are sufficiently large, the stability threshold $\varepsilon_T$ in hollow density plasmas can reach the order of magnitude of that of normal density RFP plasmas.

\begin{figure}
 \centering
  \includegraphics[width=8.0cm]{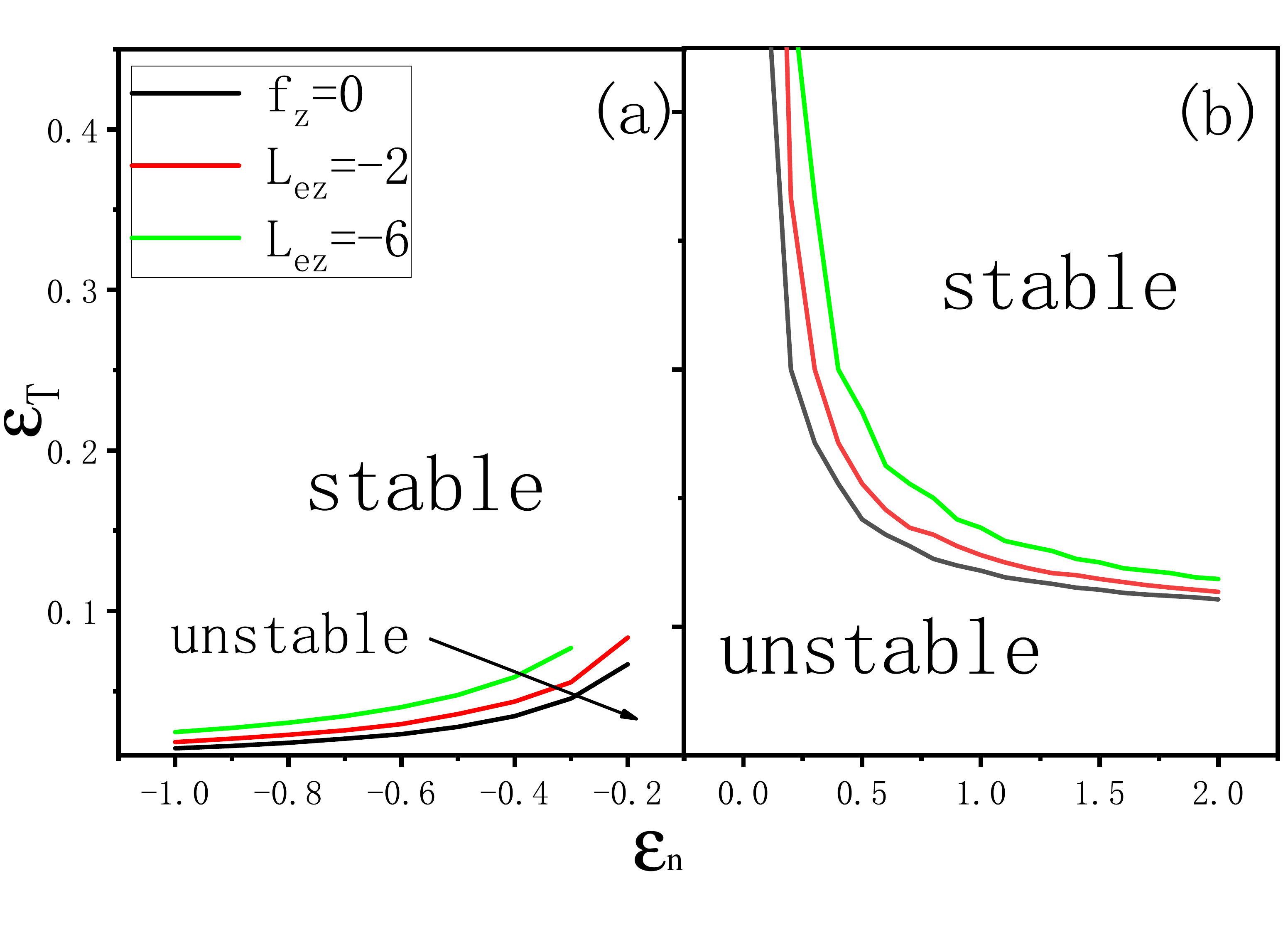}\\
  \caption{Thresholds for ITG in the ion temperature gradient ($\varepsilon_T  = L_T /R$) and density gradient ($\varepsilon_n = L_n/R$) plan in pure and mixture plasmas with hollow (a) and normal (b) density profile. Other parameters are $\hat s  = 1.0, \tau_i =\tau_z= 1.33, \eta_i= 0, \eta_z=16, k_\theta \rho_s = 0.447, \varepsilon = 0.0, q = 0.15, z = 6$ (carbon).}\label{fig:fig7}
\end{figure}

\subsection{TEMs}
In this section, we focus on the TEM instabilities in RFP plasmas with a hollow density profile and compare the results with those obtained for normal density profile plasmas. Unless otherwise stated, the parameters we adopted are as follows: $\hat s = 0.5, \eta_i=0, \tau_i = 1.33, k_\theta \rho_s = 0.447, \varepsilon = 0.15$, and $q = 0.15$. These parameters are typical characteristic parameters of the TEM excitation in RFP plasmas.

%\subsection{TEM instability in RFP plasmas with normal and hollow density profile}
The TEMs in pure RFP plasmas are discussed first. Fig. 7 depicts the normalized growth rate $\gamma/|\omega_{\star e}|$ and real frequency $\omega/|\omega_{\star e}|$ of the TEMs as a function of $\eta_e$ for the normal density profile (a)(b) and hollow density profile (c)(d) cases. This figure illustrates that $\eta_e$ provides the driving force for the TEM excitation. For the hollow density case, the TEM growth rate varies with $\eta_e$, which is similar to the non-hollow density case. Similar to the ITG mode, the difference between the hollow and non-hollow density profiles lies in the fact that the growth rates of TEM for the hollow density profile plasma are smaller. Indeed, the growth rate is around half of that observed for the non-hollow density profile plasma at the maximum growth rate. The result of these calculations indicates that the TEM instability in RFP plasmas with a hollow density profile requires a very steep density profile compared with the case of plasmas with a non-hollow density profile.

\begin{figure}
 \centering
  \includegraphics[width=8.6cm]{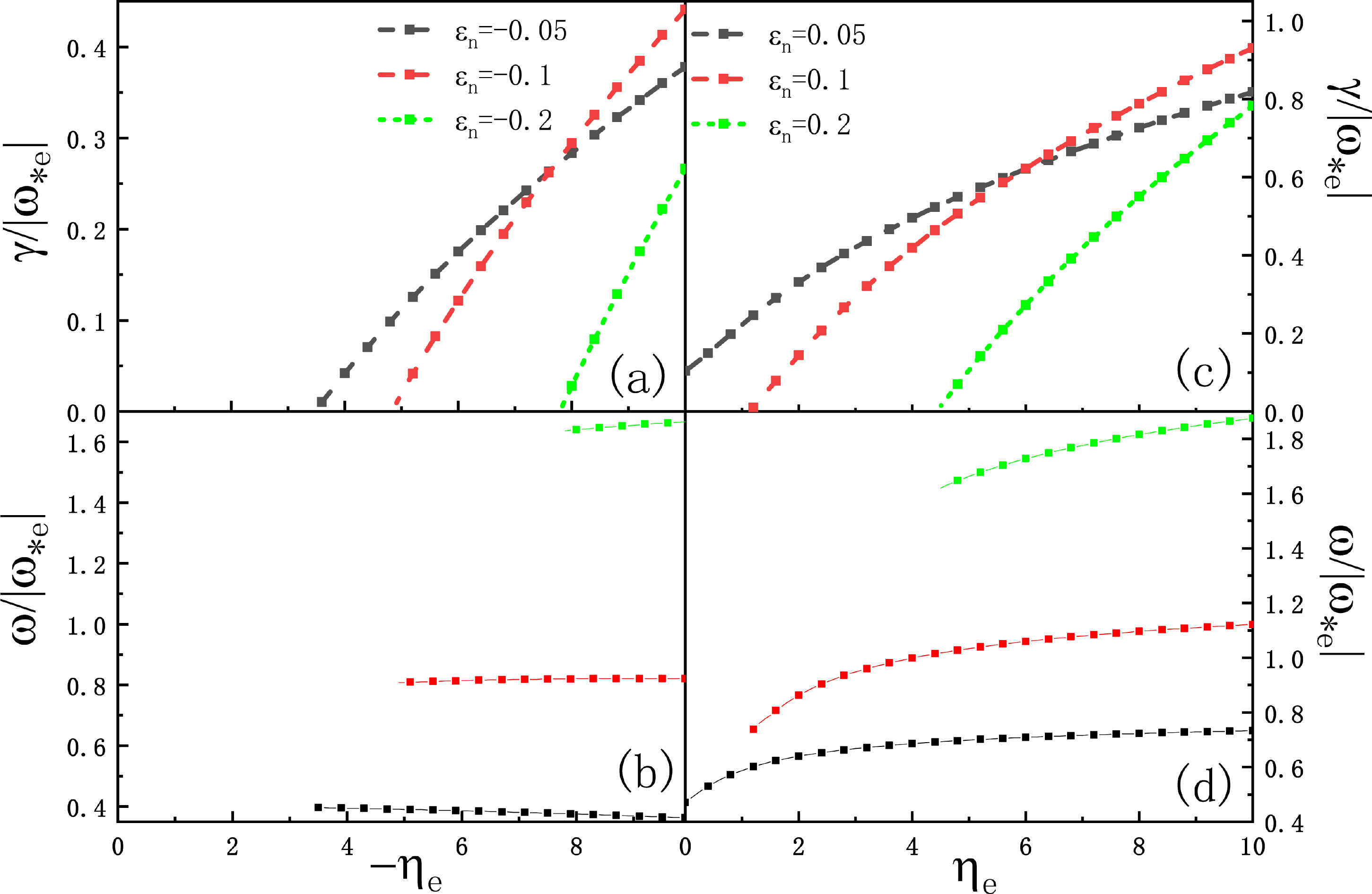}\\
  \caption{Normalized growth rate $\gamma/|\omega_{\star e}|$ and real frequency $\omega/|\omega_{\star e}|$ of TEM versus $\eta_e$ under normal density profile (a)(b) and hollow density profile(c)(d). Other parameters are $\hat s  = 0.5, \tau_i = 1.33, k_\theta \rho_s = 0.447, \varepsilon = 0.15, q = 0.15$}\label{fig:fig7}
\end{figure}

%\subsection{Stability threshold}
We performed a threshold study of the TEMs to understand at which parameters the instability arises. Fig. 8 illustrates the comparison of the stability thresholds for the TEMs in the electron temperature gradient ($\varepsilon_T$ = $L_T /R$) and density gradient ($\varepsilon_n$ = $L_n/R$) planes between the normal and hollow density plasmas. It can be clearly seen from Fig. 8 that for the normal density gradient case, the region far from the stable zone is where the electron temperature gradient and density gradient are both large. A higher temperature gradient is still required to excite the TEM instability for the hollow density case, in which the electron density gradient is larger. These results suggest that the TEM excitation occurs in a much narrow $\varepsilon_n$ region at a certain $\varepsilon_T$.

\begin{figure}
 \centering
  \includegraphics[width=8.4cm]{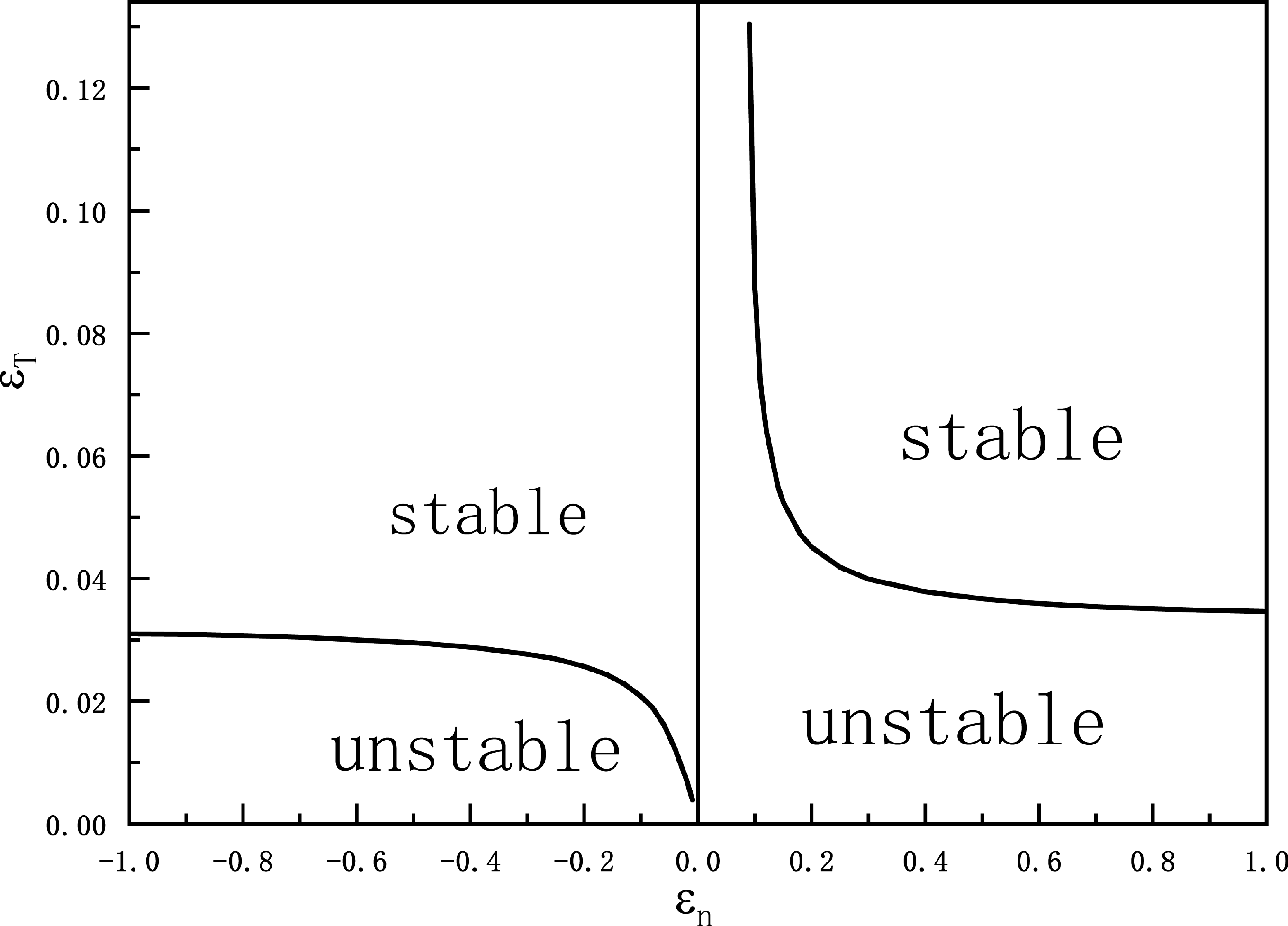}\\
  \caption{Comparison of thresholds for TEMs in the electron temperature gradient ($\varepsilon_T = L_T /R$) and density gradient
($\varepsilon_n = L_n/R$) plan between normal and hollow density plasmas. The other parameters are the same as those used in figure 7.}\label{fig:fig7}
\end{figure}

%\subsection{Impurity effects}
We carried out investigations of the impurity effects on the TEMs. As mentioned earlier, compared with the non-hollow density case, the predominant difference in the TEMs for the hollow density plasma is that their stable interval is smaller, and they are not easy to excite. Here, we mainly concentrate on the impact of impurity ions on the TEMs for the hollow density plasma. Fig. 9 depicts the normalized growth rate $\gamma/|\omega_{\star e}|$ and real frequency $\omega/|\omega_{\star e}|$ of TEMs as a function of $L_{ez}$ for the hollow density profile case; carbon $C^{+6}$ is taken as the impurity. In the presence of impurities, the relationship between the change in the TEMs and the change in $L_{ez}$ is similar to that of the ITG with $L_{ez}$; that is, as $L_{ez}$ increases (from negative to positive), the growth rate of the TEMs decreases. Note that there is a threshold value for $L_{ez}$. We refer to it as $L_{ezt}$. When $L_{ez}$ is less than $L_{ezt}$, the impurities destabilize the TEMs, while when $L_{ez}$ is greater than this value, the impurities stabilize the TEMs. For the three $f_z$ cases investigated here, the corresponding $L_{ezt}$ is around $-1$.

\begin{figure}
 \centering
  \includegraphics[width=6.3cm]{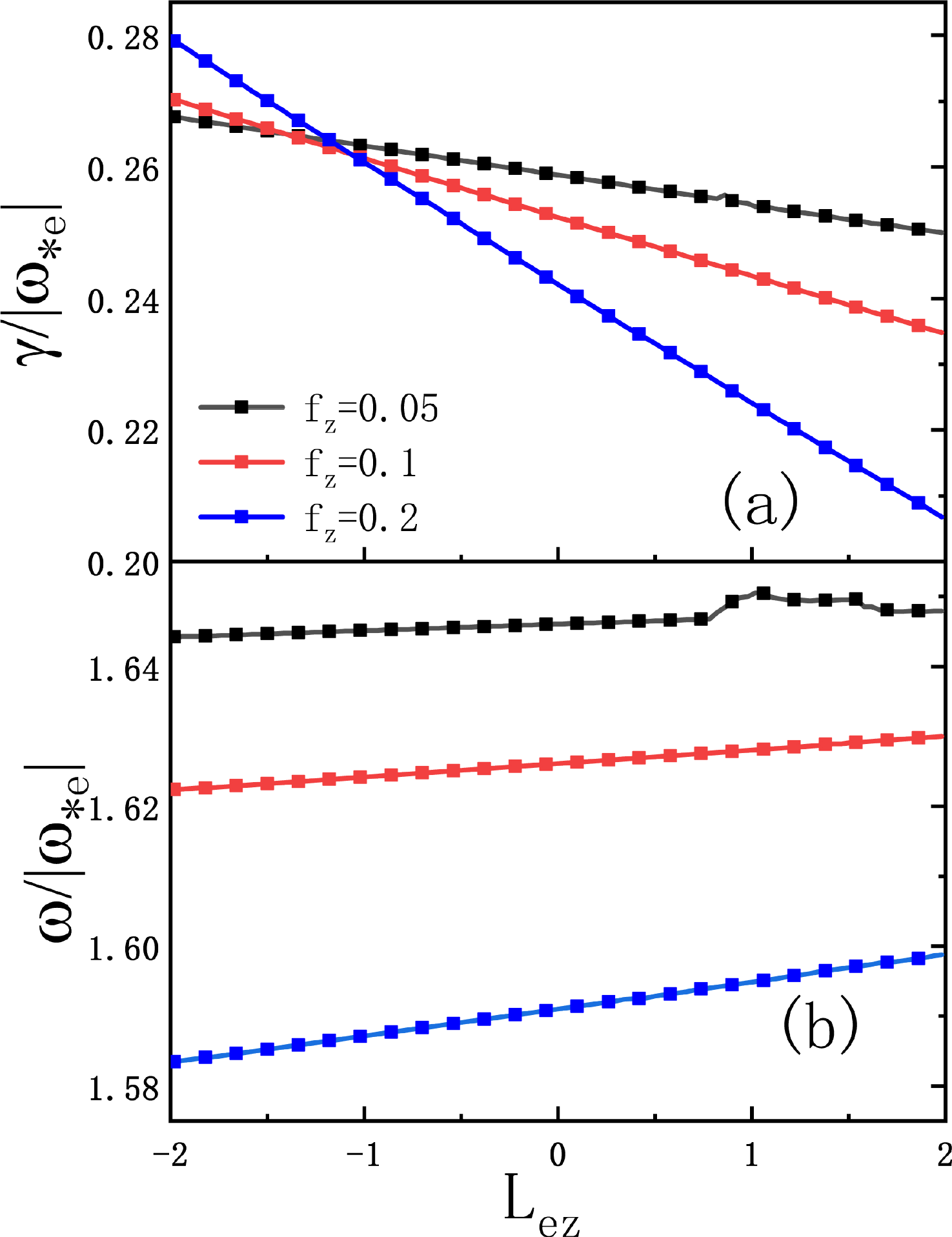}\\
  \caption{Normalized growth rate $\gamma/|\omega_{\star e}|$ (a) and real frequency $\omega/|\omega_{\star e}|$ (b) of TEM versus $L_{ez}$ under hollow density profile with different impurity charge concentration. Other parameters are $\hat s  = 0.5, \tau_i =\tau_z= 1.33, \eta_i= 0, \eta_e=10, k_\theta \rho_s = 0.447, \varepsilon = 0.15, q = 0.15, z = 6$ (carbon).}\label{fig:fig10}
\end{figure}

We then studied the effect of different impurity species. Fig. 10 shows the normalized growth rate $\gamma/|\omega_{\star e}|$ and real frequency of the TEMs as a function of $L_{ez}$ for the hollow density profile case with different impurities (namely, $C^{+6}$, $O^{+6}$, and $W^{+6}$, $W^{+12}$, and $W^{+18}$ with $f_z=0.1$). It can be seen that, for the normal density profile case and for the three lighter ions, namely $C^{+6}$, and $O^{+8}$, as $L_{ez}$ increases, the TEM growth rate decreases, while the opposite occurs for the heavy-ion $W^{+6}$. Besides, there is a mild dependence of the real frequency on $L_{ez}$ for $C^{+6}$ and $O^{+6}$; on the other hand, when the $W^{+6}$ ion is used as the impurity, the real frequency increases with $L_{ez}$. In general, both the $C^{+6}$ and $O^{+8}$ impurities reduce the growth rate of the TEMs. This is due to the fact that the finite Larmor radius of larger impurity ions can suppress the TEM instability. In addition, similar to the ITG mode, also the heavy tungsten impurity exhibits a particular characteristic. For W impurities with different charge numbers,%We consider the tungsten impurity for a more detailed analysis. Fig. 11 shows the normalized growth rate and real frequency of the TEMs as a function of $L_{ez}$ for the hollow density profile case (Figs. 6(b,d)) with W impurities ($f_z$ = 0.1) with different charge numbers, namely, $W^{+6}$, $W^{+12}$, and $W^{+18}$.
the growth rate of the TEMs decreases with the increase in $L_{ez}$, and for $W^{+6}$, the growth rate of the TEMs increases with the increase in $L_{ez}$. However, tungsten has a more pronounced stabilizing effect on the TEMs in comparison with that of other light impurities. This difference may be due to the fact that heavier impurity ions have a smaller thermal velocity; thus, the fluid condition ($\omega\gg k_{//} v_{//}$) is easily satisfied, which will weaken the wave--particle resonance effect, leading to more stable TEMs. Therefore, we can say that heavy ions with the same ionization state and different masses can easily suppress the TEM instability. In this sense, tungsten is indeed an ideal first-wall material.

\begin{figure}
 \centering
  \includegraphics[width=6.0cm]{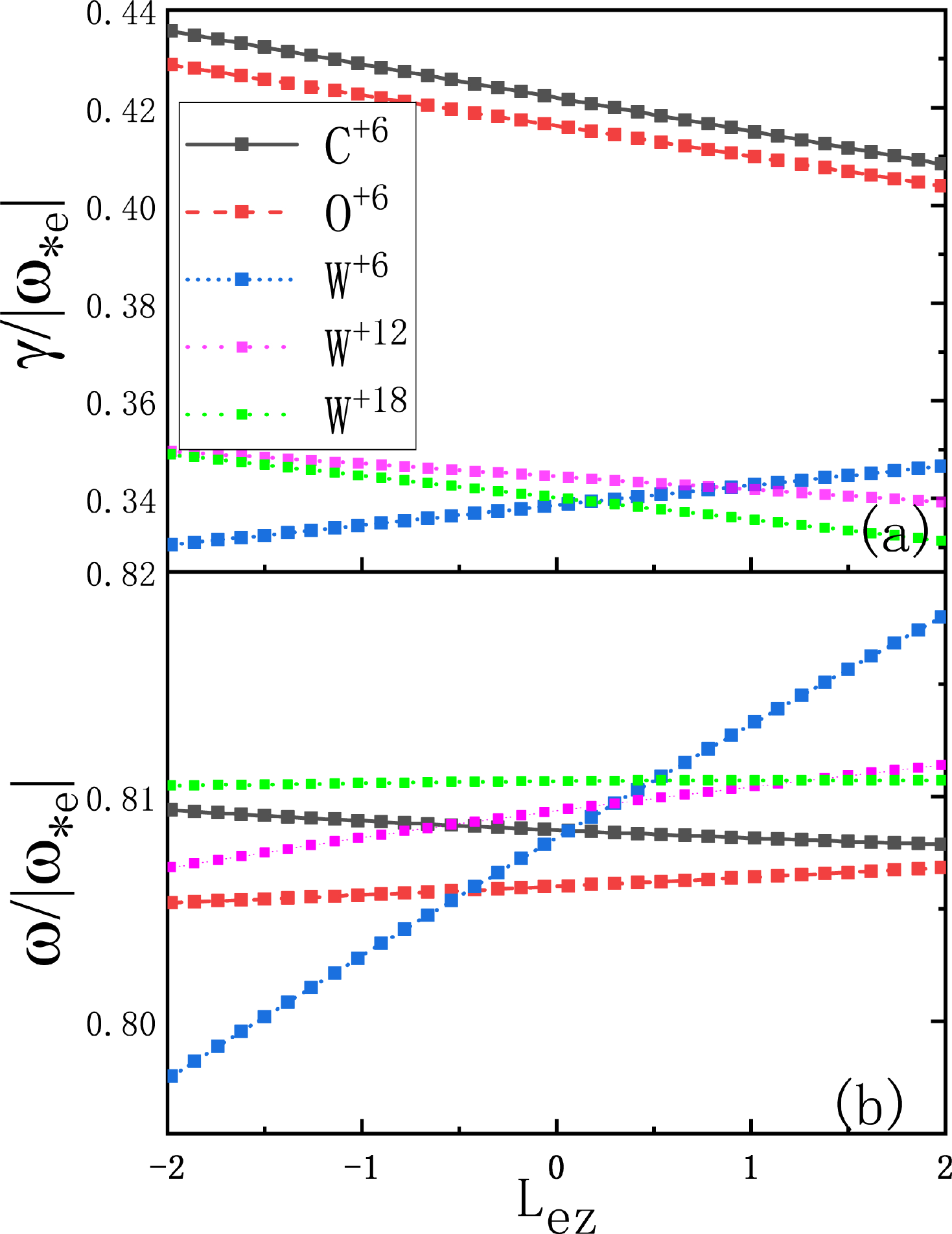}\\
  \caption{Normalized growth rate $\gamma/|\omega_{\star e}|$ (a) and real frequency $\omega/|\omega_{\star e}|$ (b) of TEM versus $L_{ez}$ under hollow density profile with different impurities($f_z=0.1$). Other parameters are the same as those used in figure 9.}\label{fig:fig10}
\end{figure}

%\begin{figure}
% \centering
%  \includegraphics[width=8.90cm]{fig12.PDF}\\
%  \caption{Normalized growth rate $\gamma/|\omega_{\star e}|$ versus $L_{ez}$ under normal (a) and hollow (b) density profile. Fully ionized carbon ($W^{+6}$, $W^{+12}$, and $W^{+18}$ is taken as impurity species with $f_z=0.1$. Other parameters are the same as those used in figure 9.}\label{fig:fig10}
%\end{figure}

The effect of the collisionality on the trapped electron mode in RFP plasmas was finally investigated. Fig. 11 depicts the normalized growth rate $\gamma/|\omega_{\star e}|$ and real frequency $\omega/|\omega_{\star e}|$ of TEMs as a function of the collisionality for both the pure and carbon RFP plasmas. Comparing the growth rates in Figs. 11(a) and 11(c), we find that when there are impurities, the growth rates for both the hollow and non-hollow density distributions are significantly lower than those obtained in the cases without impurities. At the same time, for a hollow density profile, the growth rate of the TEMs is considerably less affected by impurity ions. On the other hand, due to the impact of the collision rate, the TEM growth rate will decrease. This is because the trapped electrons can collide without completing a "banana" orbit, and the collision affects the TEM driving source. Therefore, the TEMs are damped by the collisionality.

\begin{figure}
 \centering
  \includegraphics[width=8.6cm]{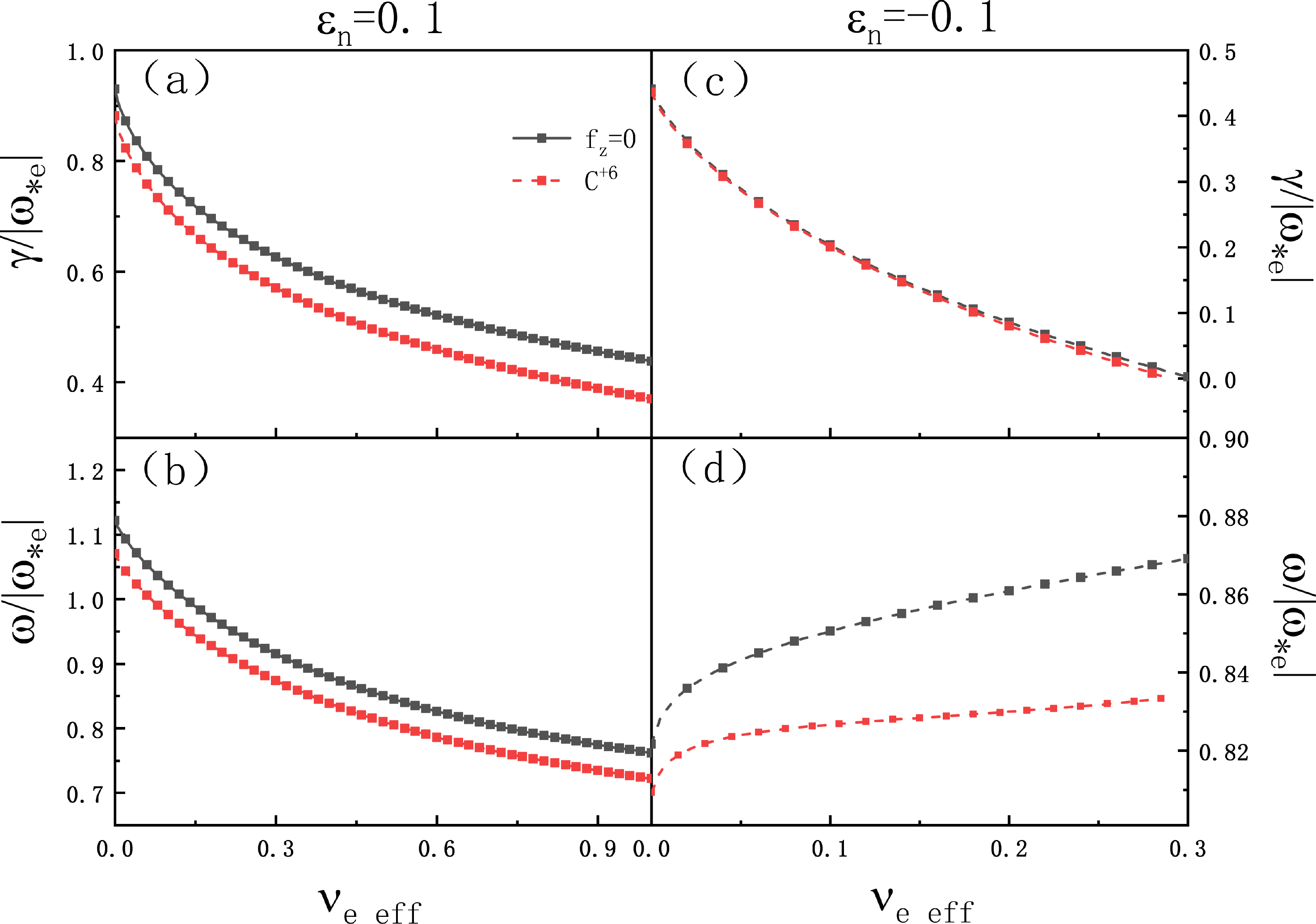}\\
  \caption{Real frequency $\omega/|\omega_{\star e}|$ and growth rate $\gamma/|\omega_{\star e}|$ versus collisionality for pure plasma ($f_z = 0$) and plasma in presence of carbon impurity ($f_z = 0.1$, and $L_{ez}=2$). The other parameters are the same as those used in figure 9.}\label{fig:fig10}
\end{figure}

%\subsection{$\eta_i$ effect}

For the TEMs, previous studies have assumed that the temperature gradient of the ions is flat. However, the realistic non-flat profile may affect the TEMs in hollow and non-hollow density plasmas. Fig. 12 shows the normalized growth rate $\gamma/|\omega_{\star e}|$ and real frequency $\omega/|\omega_{\star e}|$ of the TEMs as a function of $\eta_i$ in collisionless RFP plasmas. We see from Figs. 12(a) and 12(c) that, in the cases of hollow and non-hollow density plasmas, the growth rate decreases substantially with the increase in $\eta_i$. Hence, $\eta_i$ has a significant inhibitory effect on the TEMs. From Figs. 12(b) and 12(d), one also finds that $\eta_i$ has a little effect on the real frequency of the TEMs, especially in the case of hollow density plasmas. In addition, the increase in $\varepsilon_n$ also leads to a decrease in the TEM growth rate. For example, $\varepsilon_n=2$ will cause the TEMs to exist only in a small range of $\eta_i$ values. Nevertheless, $\varepsilon_n$ will cause an increase in the real frequency, even though this larger frequency only exists in a small range of $\eta_i$ values.

Our results show that the instability interval of the TEM mode in RFP plasmas is much smaller than that of tokamak plasmas. For example, when the plasma density gradient is small, a high electron temperature gradient is required to excite the TEMs. The main reason for this is that there is a higher ion Landau damping in RFP plasmas than in tokamak plasmas (the lower $q$ in RFP plasmas will result in a more significant Landau damping [26]). Our simulations show that the Landau damping can not only suppress the ITG mode; that is, as the Landau damping increases, the free energy that drives the ITG mode is reduced, and the ITG mode becomes more stable. Indeed, the Landau damping can also stabilize the TEMs. We found that when the effective ion Landau damping frequency changes with respect to the electron precession resonance by increasing the temperature ratio $\tau$, that is, reducing the ion Landau damping, the growth rate of the TEMs and the TEM stable interval increase. Therefore, the high Landau damping in RFP plasmas makes it difficult to excite the TEMs.

\begin{figure}
 \centering
  \includegraphics[width=8.6cm]{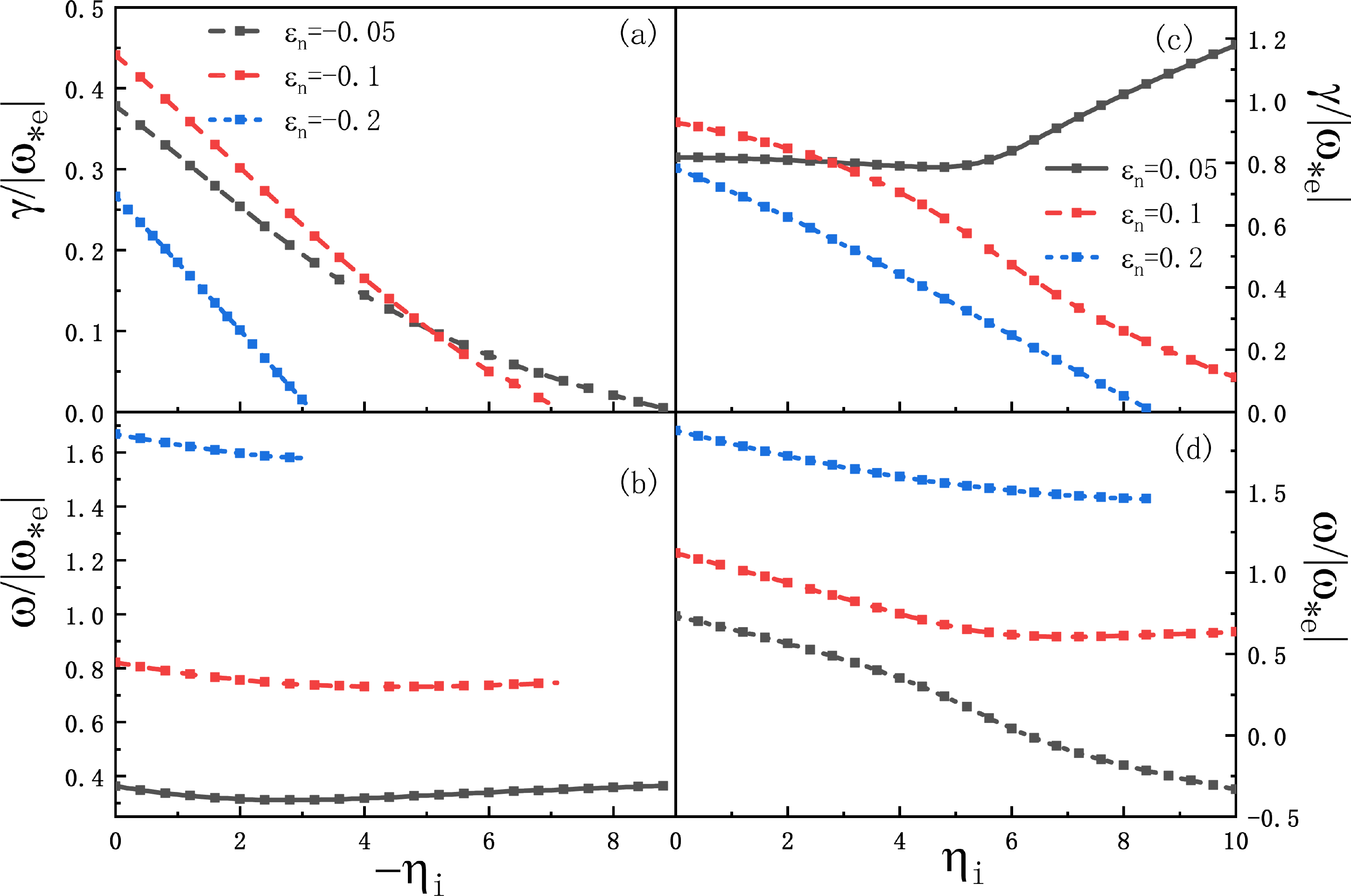}\\
  \caption{ Real frequency $\omega/|\omega_{\star e}|$ and growth rate $\gamma/|\omega_{\star e}|$versus $\eta_i$. The other parameters are the same as those used in figure 9.}\label{fig:fig10}
\end{figure}

Interestingly, a reversed density profile can further stabilize the TEMs in RFP plasmas but cannot eliminate the TEMs. A possible reason is that the precession drift of the trapped electrons does not depend on the density gradient. The rotation direction of the TEMs for the hollow density profile case is changed, making the resonance effect of the trapped electrons and modes invalid. Hence, the TEM instability becomes very weak (the collisionless TEM excitation is mainly due to the resonance of the instabilities and the precession drift of the trapped electrons). These results are qualitatively consistent with those of Tang et al. [29].

\section{Summary}
\label{sec:4}

The gyrokinetic theory was used to study the impact of impurities on the drift wave instability in toroidal plasmas with regular and hollow density profiles. Both collisionless and collisional RFP plasmas with normal and hollow density profiles were considered.

The first part of the article discussed the ITG investigations. For situations with impurities, we found that when $L_{ez}$ is positive, impurities can reduce the ITG instability, while they can increase the instability when $L_{ez}$ is negative; the greater the concentration of impurities, the more pronounced this phenomenon. In addition, by comparing the results of the regular ($\varepsilon_n > 0$) and hollow ($\varepsilon_n < 0$) plasma density profiles, we found that the growth rate of the former is smaller, the instability interval is smaller, and the ITG instability is harder to excite.

In the second part, the present work focused on studying the influence of different impurities on the TEM instability. The selected impurities were fully-ionized carbon, oxygen, and tungsten. Through simulations, we found that the TEM growth rate for the hollow density case is still lower than that for the normal density case. Moreover, for the concave density profile, the change in the TEMs and $L_{ez}$ is similar to that of the ITG mode with $L_{ez}$ with impurities. That is, as $L_{ez}$ increases (from negative to positive), the growth rate of the TEMs decreases. There is a threshold value for $L_{ez}$. When $L_{ez}$ is less than this threshold, the impurities destabilize the TEMs, while when $L_{ez}$ is greater than this threshold, the impurities stabilize the TEMs. The results are the opposite in the case of W impurities. Tungsten is unique due to its high mass number. The TEMs are harder to excite when W impurities are incorporated, suggesting that W is an ideal first-wall material.

Finally, the impact of collisionality on the TEM instability was studied. The simulation results show that, considering the collisionality, the growth rate of the TEMs decreases with increasing the collision rate, which is consistent with previous calculation results. However, in the case of both the typical and hollow density profiles, as $\eta_i$ increases, the growth rate of the TEMs also decreases. We have clarified the physics underlying the TEMs. Besides, the instability threshold analysis in this article will provide a reference for comparing simulation and experimental results.

Work involving electromagnetic simulations of the ITG and TEMs in toroidal plasmas is ongoing alongside comparisons of the drift wave characteristics in both RFP and tokamak plasmas.

\section*{Acknowledgments}
\label{sec:5}
This work is supported by the National Natural Science Foundation of China (Nos. 11905109 and 11947238), the National Key $R\&D$ Program of China (Nos. 2018YFE0303102 and 2017YFE0301702), and the Center for Computational Science and Engineering of Southern University of Science and Technology. J. C. Li also thanks Y. Liu at NKU, M. Xu, and J. Q. Xu at SWIP for fruitful discussions.


\section{References}
\begin{thebibliography}{42}

 \bibitem{1}W. Horton, 1999 {\it Rev. Mod. Phys.} \textbf{71} (3), 735.

       \bibitem{2}F. Wagner, et al., 1982 {\it Phys. Rev. Lett.} \textbf{49} (19), 1408.

   %    \bibitem{3} D. Galassi, et al., 2017 {\it Nuclear Materials and Energy} \textbf{12} 953-958.
       \bibitem{3} Hongxuan Zhu, et al., 2018 {\it Phys. Rev. E} \textbf{97} 053210.

       \bibitem{4} R. A. Heinonen and P. H. Diamond, 2020 {\it Phys. Rev. E} \textbf{101} 061201(R).

        \bibitem{34} X. Garbet, 2001 {\it Plasma Phys. Control. Fusion} \textbf{43} A251.

       \bibitem{4}Z. Lin, et al., 1998 {\it Science} \textbf{281} (18), 1835.

        \bibitem{5}A. Hasegawa, 1969 {\it Phys. Fluids} \textbf{12}, 2642.

          \bibitem{6}Jingchun Li, et al., 2021 {\it Plasma Phys. Control. Fusion} \textbf{63} 125005.

          \bibitem{67}J. R. Duff, et al., 2018 {\it Physics of Plasmas} \textbf{25} 010701.

           \bibitem{677}Z. R. Williams, et al., 2017 {\it Physics of Plasmas} \textbf{24} 122309.

       \bibitem{7} C. Angioni, et al., 2017 {\it Nuclear Fusion} \textbf{57} 116053.

       \bibitem{8}M. Valovic, et al., 2008 {\it Nuclear Fusion} \textbf{48} 075006.
	
		\bibitem{9} G. A. Wurden, et al., 1987 {\it Nuclear Fusion} \textbf{27} 857.
		
		\bibitem{10} B. Baiocchi, et al., 2015 {\it Nuclear Fusion} \textbf{55} 123001.

       \bibitem{11} M. Zuin, et al., 2017 {\it Nuclear Fusion} \textbf{57} 102012.
		
	  	\bibitem{12} T Barbui, et al., 2015 {\it Plasma Phys. Control. Fusion} \textbf{57} 025006.
		
	 	\bibitem{13} B Pegourie, 2007 {\it Plasma Phys. Control. Fusion} \textbf{49} R87.
		
		\bibitem{14} X. Ding, et al., 2006 {\it Chinese Physics Letters} \textbf{23} 2502.

   \bibitem{15} R. Lorenzini, et al., 2002 {\it plasma Phys. Control. Fusion} \textbf{44} 233.

		\bibitem{16} R. Sakamoto, et al., 2006 {\it Nuclear Fusion} \textbf{46} 884.
		
		\bibitem{17} L. Garzotti, et al., 2014 {\it Plasma Physics and Controlled Fusion} \textbf{56} 035004.
		
		\bibitem{18} J. Q. Dong, et al., 2001 {\it Physics of Plasmas} \textbf{8} 3635-3644.
		
		\bibitem{19} Jingchun Li, et al., 2020 {\it Plasma science and Technology} \textbf{22} 055101.
		
		\bibitem{20} I. Predebon, et al., 2011 {\it plasma Phys. Control. Fusion} \textbf{53} 125009.
		
		\bibitem{21} J.C. Li, et al., 2019 {\it EPL (Europhysics Letters)} \textbf{127} 45002.
		
		\bibitem{22}S. C. Guo, {\it Physics of Plasmas} \textbf{15} 122510.
		
		\bibitem{23} S. F. Liu, et al., 2014 {\it Nuclear Fusion} \textbf{54} 043006.

       \bibitem{24} S. F. Liu, et al., 2011 {\it Nuclear Fusion} \textbf{51} 083021.

    %20    \bibitem{15} J. B. Taylor, et al Plasma Physics,\textbf{10} 1835(2010) .
		
%	21	\bibitem{16} F. Romanelli, et al,Physics of Fluids B: Plasma Physics \textbf{5} 4081(1993).
		
		
		%\bibitem{18} Liu, Feng  and  Lin, Z.  and  Dong, J. Q.  and  Zhao, K. J,Physics of Plasmas,\textbf{17} (2010) 1835.
		
		
		
	%	\bibitem{20} Dong, J. Q.  and  Horton, W.  and  Dorland, W.,Physics of Plasmas,\textbf{1} (1994) 3635-3640.



      \bibitem{25}W. M. Tang, et al., 1975 {\it Physical Review Letters} \textbf{35}(10) 660.


\end{thebibliography}
\end{document}